# Possible Atmospheric Transparency Studies on the Basis of Cherenkov Light Measurements

*Mishev A.L.[1]*

[1] *Nuclear Regulatory Agency, Sofia 1574, Bulgaria, alex_mishev@yahoo.com*

**Abstract:**
The possibility to use Cherenkov light measurements with Cherenkov telescope to study atmospheric processes is shown.
Cherenkov light from extensive air showers is obtained using Monte Carlo simulations with CORSIKA code. Different atmospheric profiles are considered and compared.
Several experimental results are shown and the scientific potential is discussed.

## 1. Introduction

Primary cosmic rays (CR) penetrate Earth atmosphere and depose energy via nuclear interaction and ionization losses. Two components of CR are important for cosmic ray atmosphere interactions - galactic cosmic rays (GCR) and solar CR or solar energetic particles (SEP).

Solar CRs are accelerated during the explosive energy release at the Sun and by acceleration processes in the interplanetary space. Solar CRs enter the atmosphere sporadically, with a higher probability during periods of high solar activity.

The SEPs are in positive correlation with ongoing solar activity while GCRs are in anticorrelation with the main solar activity indices.

The effect of cosmic rays on climate changes is widely discussed. In several studies is suggested that cosmic rays (CR)s, namely their variations are factor leading to climate changes trough large diversity of mechanisms [1-4]. Obviously the Sun is the main climate driver. However CRs can play significant role, because the measured solar irradiance variability is small to explain the observed climate variations.

The solar variability is connected with indirect mechanism via CR to climate change such as atmospheric ionization. The CR induced ionization is the main ionization source in low and middle atmosphere and is related to cloud formation [5-6] trough cosmic ray-aerosol-cloud interactions [7-8].

Obviously the change of cloud cover leads to atmospheric transparency changes. Therefore the transparency, which is one of the primary measures of the atmospheric state, is connected with cosmic ray, respectively solar variations.

In this paper is demonstrated the possibility to study atmospheric transparency using Cherenkov light flux measurements [9].

## 2. Atmospheric Cherenkov light and extinction

The atmospheric Cherenkov light is produced by charged ultra relativistic particles in extensive air showers (EAS).

The majority of the Cherenkov photons are produced near to the shower maximum [10]. As a result the quasi totality of the Cherenkov light passes trough the lower atmosphere.

The generation and propagation of Cherenkov radiation is affected by atmospheric conditions [11].

The Cherenkov light undergoes extinction in the atmosphere because of absorption on molecules by Rayleigh scattering and Mie scattering by aerosols. Therefore measuring atmospheric Cherenkov light produced by cosmic ray it is possible to estimate the atmospheric transparency Fig.1.

In addition it is claimed that Cherenkov light absorption in the atmosphere is negligible and only molecular and aerosol scattering of photons is taken into account [11].

Molecular scattering is almost constant, whereas aerosol concentration in the boundary layer above the surrounding terrain is of diurnal and seasonal variability.

The atmospheric extinction coefficient may be evaluated using Cherenkov light measurements and following procedure described in [12].

The method is based on the fact that the Cherenkov light flux is proportional to the primary energy: Therefore the extinction is defined [12]

$$\varepsilon_i = \varepsilon_0 \left( \frac{N_i(>Q_{thr})}{N_o(>Q_{thr})} \right)^{1/\kappa} \tag{1}$$

where $N(>Q_{thr})$ is the integral number of events with the light intensity above the threshold detected in the $i^{th}$ and basic periods; $\varepsilon$ is the atmospheric extinction coefficient; $\kappa$ is the energy spectrum index.

The resultant extinction coefficient is a product of the variable $\varepsilon$ due to the current aerosol concentration and the basic coefficient $\varepsilon_0$ caused by the molecular scattering and minimal aerosol extinction.

It is possible to evaluate $\varepsilon_0$ as function of $X_{max}$ as a parabolic function and thus obtain the requested extinction.

Atmospheric extinction varies with location and altitude. The amount of atmospheric extinction depends on the altitude of an object, being lowest at the zenith and at a maximum near the horizon. It is calculated by multiplying the standard atmospheric extinction curve by the mean airmass calculated over the duration of the observation. In this connection measurements at various zenith angles are important, as cross calibration with other methods such LIDAR measurements or classical astronomical measurements.

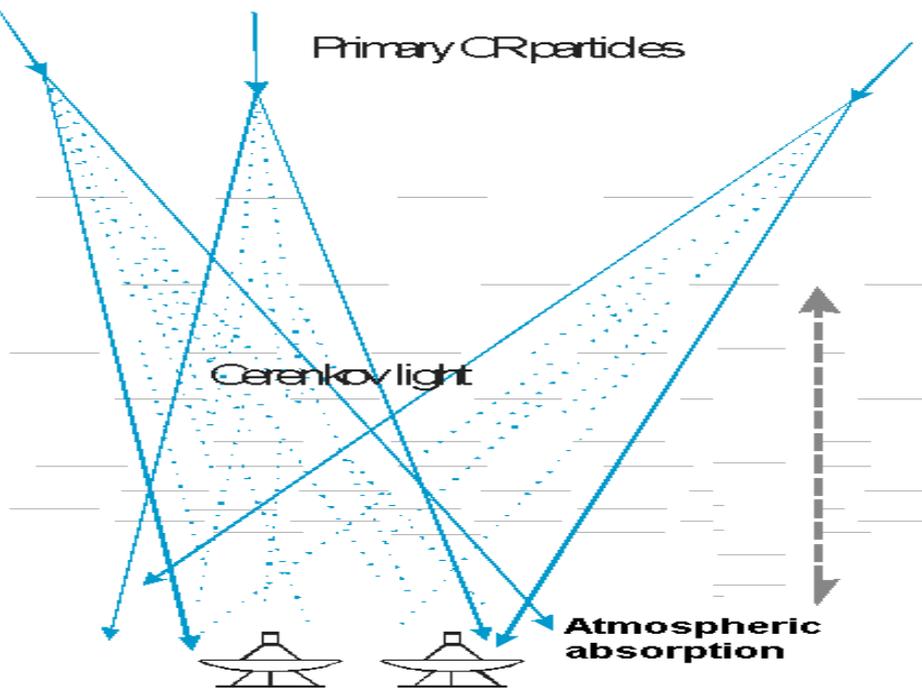

Fig.1.Estimation of atmospheric transparency on the basis of Cherenkov light registration

*2.1 Seasonal and diurnal variations*

Atmospheric density profiles depend on geographic position and are generally time-variable. Several light absorption and scattering processes are connected with atmospheric density. Different density profiles lead to differences in Cherenkov light density of up to 60%. Seasonal variations at mid-latitude sites are of the order of 15–20% [11].

Moreover the amount of production of Cherenkov photons in the atmosphere, as well their lateral distribution depends on atmospheric depth. As example the contribution of Raleigh attenuation and Mie scattering at observation level of 5000 m above sea level is practically negligible Fig.2. The presented in Fig. 2 results are obtained on the basis of Monte Carlo simulations with CORSIKA 6.52 [13] code using FLUKA [14] and QGSJET II [15] hadronic interaction models. The mentioned above effects of scattering and absorption of Cherenkov light take place at lower observation levels.

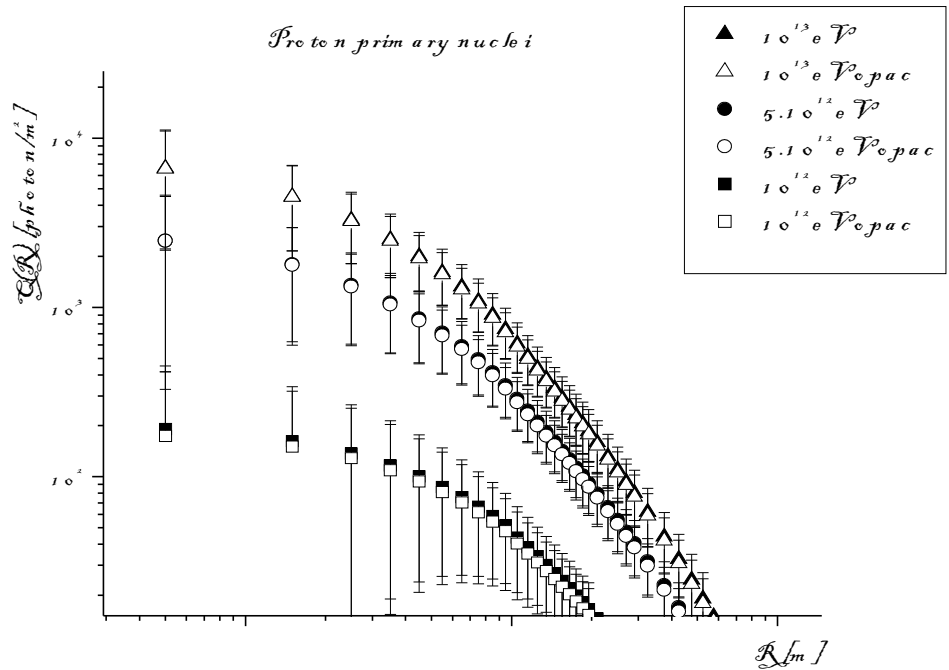

Fig.2. Lateral distribution of Cherenkov light with and without Raleigh attenuation and Mie scattering

Thus a detailed modeling of Cherenkov light production as a function of season, geographic place and altitude is necessary in attempt to admit such effects.

In this connection the effect of CR variation on atmosphere, namely transparency in daily/weekly time scales seems possible.

It is known that regular daily variations of CR flux are in the order of 1%. In addition some transient phenomena occur, which can reduce or enhance CR flux.

As example the variations in atmospheric transmission of several percent in clear air, accompany solar wind events associated with variations on the day-to-day timescale in the flow of vertical current density in the global electric circuit. These events occurred when there was a high loading of stratospheric aerosols.

Decreases in transmission, are present when Forbush decreases of galactic cosmic ray flux occur, but only during periods of low stratospheric aerosol loading. Forbush decreases are associated with both tropospheric ion production decreases and current density decreases.

Similar effects are present on the 11-year solar cycle, with climate consequences that have yet to be analyzed [16]. The mechanisms for these phenomena are not well understood, but the nature of the observations suggests that explanations should be sought in terms of theories of the effects of electric charge on the formation of aerosols.

In addition the variations of solar and galactic cosmic rays may be responsible for the changes in the large-scale atmospheric circulation. It is possible to associate such type of phenomena with solar activity, precisely with cosmic particles of 0.1–1 GeV [17]. Possible mechanism of cosmic ray effects on the lower atmosphere involves changes in the atmospheric transparency, which is connected with cloud cover. This is due to changes in the stratospheric ionization produced by the considered cosmic particles, during the solar cosmic ray bursts [18].

Therefore the measurements of atmospheric Cherenkov light can contribute significantly to studies related to impact of CR on Earth's atmosphere on daily/weekly time scales and for some case studies.

*2.2 Annual variability*

The 11-year solar cycle is the main source of periodic solar variations. He is governed by a hydromagnetic dynamo process. The cycle reflects on structures the Sun's atmosphere, corona and wind. During the cycle modulations of the solar irradiance, frequency of flares, coronal mass ejections, and other geoeffective solar eruptive phenomena, and the flux of high-energy galactic cosmic rays entering the solar system occur.

The expansion of solar ejections into interplanetary space provides overdensities of plasma that are efficient at scattering high-energy cosmic rays entering the solar system. Since the frequency of solar eruptive events is strongly modulated by the solar cycle, the degree of cosmic ray scattering in the outer solar system varies. As a result, the CR flux in the inner solar system is in anticorrelation with the level of solar activity.

In this connection the CR-climate connections are mostly studied on annual time scale. A decadal cycle in global cloud coverage was reported [3, 5]. This hypothesis gave rise to big discussion pros and cons. Several recent studies [19] demonstrate clear direct correlation between low cloud cover and CRs in few regions, which roughly corresponds to model predictions [20].

In addition a significant correlation between low cloud coverage and cosmic rays was observed during last 22 years [21] on the basis of International Satellite Cloud Climatology Project data. The peak positive correlation is observed at 50 degrees North and South with a tendency to negative correlation at lower latitudes. The correlation is strongest over the North and South Atlantic.

In this connection the study based on atmospheric Cherenkov light measurements seems very promising.

## 3. Preliminary Results

Usually the instrument, which provides information about the status of the atmosphere is a lidar. Such type of device is rather expensive. At previous section was mentioned the possibility and scientific potential to apply atmospheric Cherenkov technique for atmospheric studies. At the same time it exist several Cherenkov telescopes in operation.

A simple two-mirror telescope could be used for atmospheric Cherenkov light registration. Measuring the Cherenkov light flux produced in EAS in different atmospheric conditions it is possible to obtain different amplitude spectra. The different slopes of the reconstructed spectra correspond to different atmospheric conditions [9, 22]. In addition calibrated instrument measurements gives the possibility to obtain directly the extinction following [12] and using (1).

The used for measurements device is the developed Cherenkov telescope at Basic Environmental Observatory Moussala, which represents system of two parabolic mirrors with focal length of 1.5m and diameter of 2m, working in a coincidence regime Fig. 3a,b.

A detailed Monte Carlo simulation of the detector response at different atmospheric conditions was carried out. For this purpose CORSIKA 6.52 [13] code with FLUKA [14] and QGSJET II [15] hadronic interaction subroutines was used. The obtained amplitude spectra are presented in Fig. 4.

In Fig. 4 we observe similar shape of the obtained Cherenkov amplitude spectra, but with various amplitudes.

Thus taking into account the observed difference especially for low amplitudes it is possible to distinguish the atmospheric Cherenkov light flux in different atmospheric conditions. Moreover this difference is obtained using the assumption of low aerosol load in the atmosphere considering only Raleigh attenuation and Mie scattering without absorption.

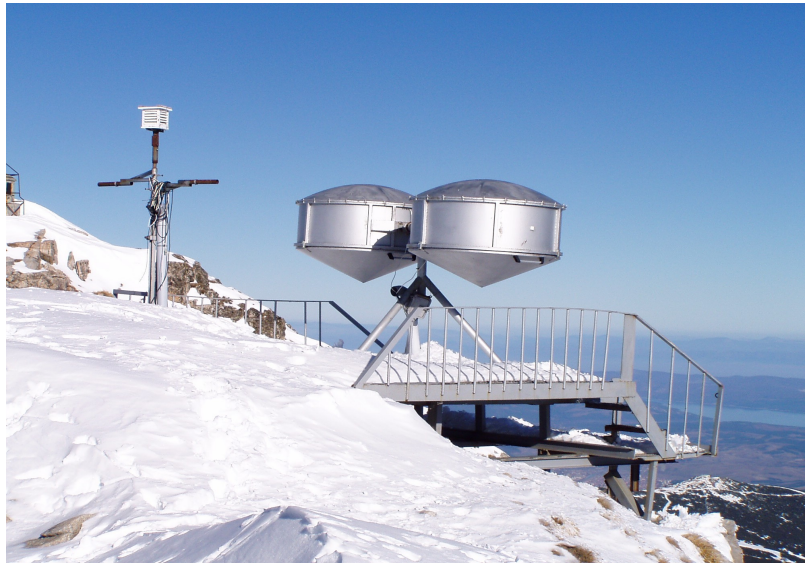

Fig.3a. The Cherenkov telescope at BEO Moussala

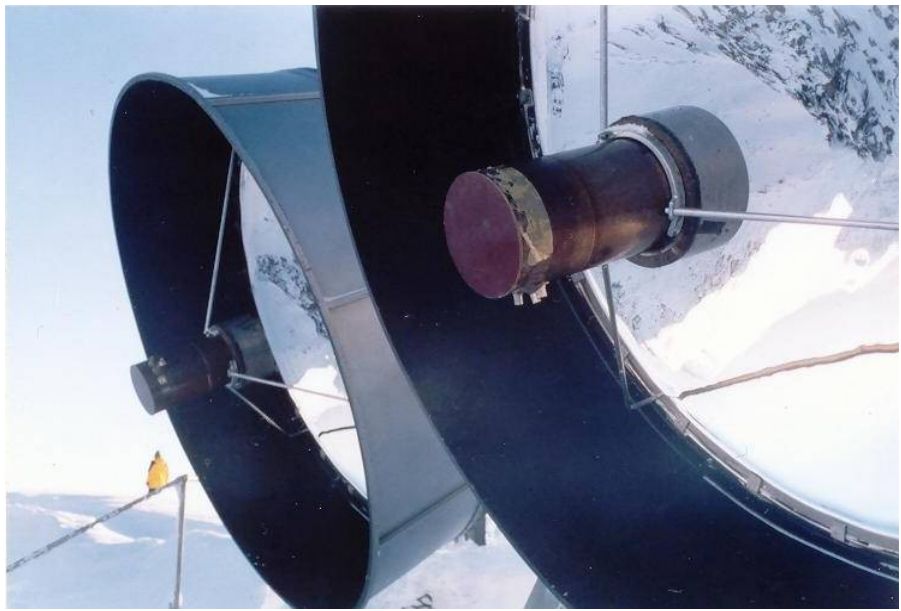

Fig. 3b The Cherenkov telescope at BEO Moussala

Several preliminary measurements carried out with experimental setup similar to BEO Moussala Cherenkov telescope at sea level observation level confirmed the expectations [9, 22, 23] Fig. 6.

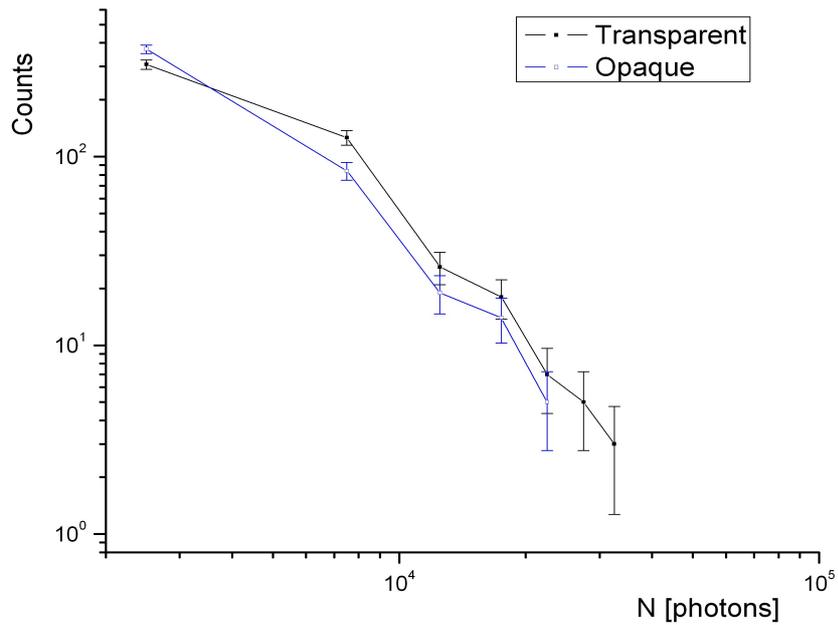

Fig.4.Simulated amplitude Cherenkov spectra at different atmospheric conditions

In addition several measurements of aerosol optical thickness are carried out with MICROTOPS II ozonometer are carried out. The aim is to obtain information concerning atmospheric transparency during daytime.

This is a 5 channel Fig. 5 hand-held device for measuring total ozone column and give an additional information about total water vapour and aerosol optical thickness, developed by SolarLight.

The cross correlation analysis is performed using the data of mentioned above devices.

Even the lack of statistics a promising cross correlation coefficient of 0.57 is obtained for

Cherenkov and cosmic ray variation data and 0.47 for aerosol optical depth and cosmic ray variation data.

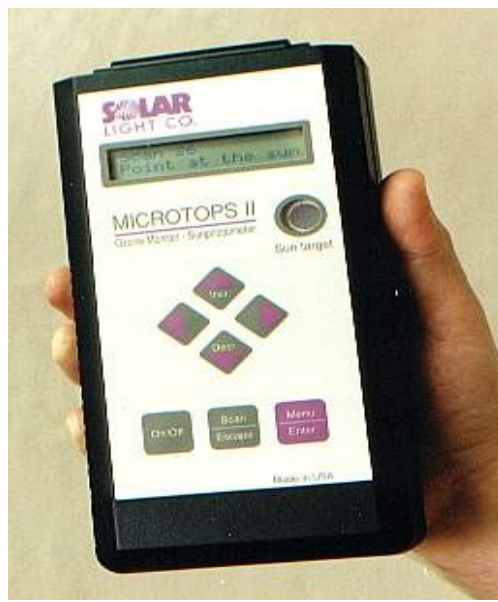

Fig. 5 MICROTOPS II ozonometer

## 4. Discussion

The atmospheric transparency is one of the primary indications related to atmospheric state. The precise long term series of atmospheric transparency measurements gives the possibility for quantitative estimate of the variability of air circulation and make climatologic conclusions with regard to contamination, cloud formation, humidity and radiative exchange.

Atmospheric transparency varies with location, altitude and season.

In addition the transparency is connected with atmospheric turbidity [24]. As example diurnal and annual variations of the atmospheric turbidity are found, with a summer afternoon maximum and a winter morning minimum. In this case the correlation between atmospheric turbidity and specific humidity shows that the summer maximum is due to the heavy water vapour content of maritime air masses, carried by the west–southwestern winds prevalent during this season. Continental dust particles, carried by the east–northeastern winds, growing due to water vapours result in high turbidity at the end of summer. The winter minimum is caused by a considerable decrease of the humidity and dust content of the continental air masses, carried by strong east–northeastern winds, prevalent during the cold period.

Atmospheric extinction has three main components: Rayleigh scattering by air molecules, scattering by aerosols, and molecular absorption. The most important sources of molecular absorption are molecular oxygen and ozone, which absorb strongly in the UV, and water vapor, which absorbs strongly in the infrared. The latter is not topic of interest in this study, because Cherenkov light maximum is in the near UV. After the scattering of Cherenkov light at lower observation levels the maximum is shifted to visible region.

The amount of atmospheric extinction depends on the altidute of the object. It is lowest at the zenith and maximum near the horizon. In theory it can be estimated by multiplying the standard atmospheric extinction curve by the mean airmass calculated over the duration of the observation.

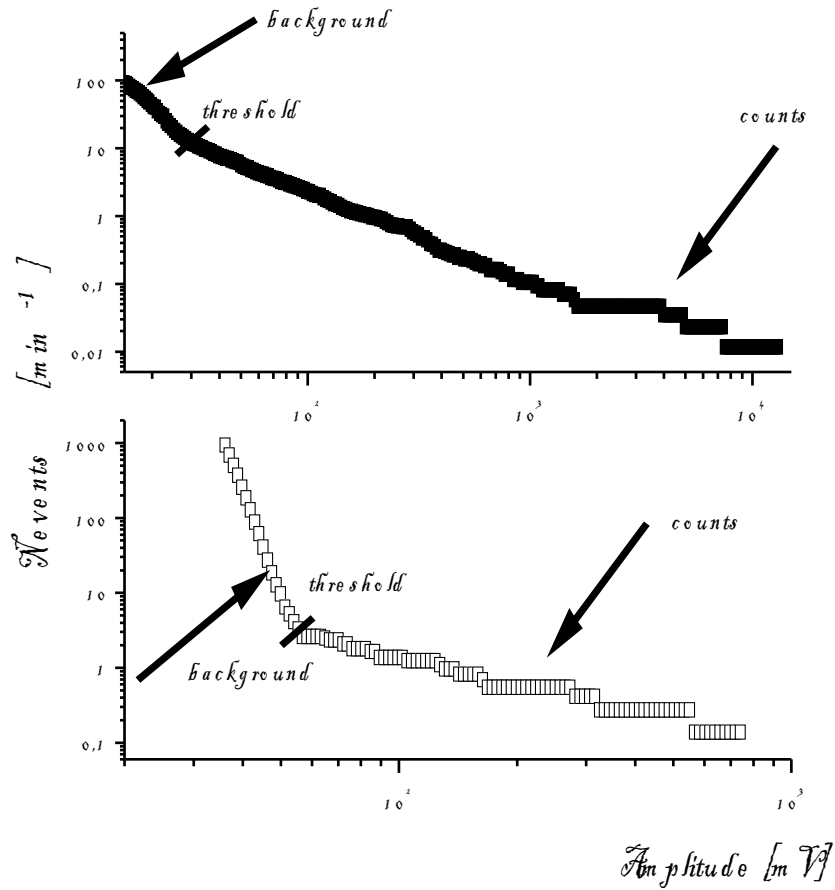

Fig.6.Measured amplitude Cherenkov spectra at various atmospheric conditions

First of all this reason we should calibrate the Cherenkov measurements for season and location. This is possible on the basis of theoretical estimations or local measurements of atmospheric extinction by astronomical observations.

Afterwards the long term Cherenkov measurements can be applied for atmospheric studies due to cosmic ray processes.

Summarizing the continues Cherenkov light measurements can be applied to study atmospheric transparency changes due cosmic ray variations and cosmic ray processes as principal aim of this study. In this connection ad additional measurements of cosmic ray variations are of big interest [9].

This will permit to study the effects of solar variability respectively cosmic ray variation on regional climate time series [25].

The second goal as was mentioned above is to investigate atmospheric changes due to sporadic CR effects such as Forbush decreases [16, 26] and associated effects due to primary cosmic ray protons especially in the range of GeV energies. Such type of studies will give good basis to check different supposed mechanisms [27].

Finally the long series of Cherenkov light measurements provide basis to investigate seasonal variations of atmospheric state and effects related to pressure changes or internal gravitational waves.

In general the atmospheric transparency studies, based on this method can contribute to understand to cosmic-ray-cloud formation relation mechanism, as well atmospheric circulation processes.

Presently an essential progress in development of physical model for cosmic ray induced ionization processes in the atmosphere was carried out [28-31]. However how cosmic ray induced ionization affect cloud formation is not well understood.

The estimations based on Monte Carlo simulations Fig.4 demonstrate that even cloudless atmosphere with Rayleigh and Mie scattering affects the expected amplitude spectra, without changing the slope [22, 23].

As was shown in Fig. 5, the measured spectra confirmed this result. The impact of cloudy atmosphere is on amplitude spectrum i.e. the presence of clouds leads to diminishing of measured amplitudes.

However the reconstructed slopes of measured spectra are with similar values for cloudy and clean atmosphere.

The possibility to estimate the atmospheric transparency on the basis of measurements of cheap and simple for exploitation device such as Cherenkov telescope contribute to fundamental understanding of processes connected with ion-aerosol-cloud formation [32].

Both proposed mechanisms connected with cosmic ray induced ionization in the atmosphere, which can be realized by ion-induced aerosol nucleation or electrically-enhanced contact nucleation are of interest for study.

In the model for ion-induced aerosol nucleation [33, 34], sulfuric acid $H_2SO_4$ and water are assumed to be the most important nucleating agents in the free troposphere, namely, negative ion-$H_2SO_4$-$H_2O$ nucleation is a source of new particles in the troposphere.

The electrically-enhanced contact nucleation [35, 36] is related to the influence of interparticle electrical forces on the particle collection rates of charged water drops.

Electrical charges on aerosol particles and droplets modify the droplet-particle collision efficiencies involved in scavenging, and the droplet-droplet and particle- particle collision efficiencies involved in coalescence of droplets and particles [37].

In general the Monte Carlos simulations, the preliminary experimental data and discussed above scientific potential show that the proposed method seems very promising.

**Acknowledgement**

The author acknowledges the colleagues the team of Prof. P. Velinov from Solar-Terrestrial Influences Institute of Bulgarian Academy of Sciences.

REFERENCES

[1] L. Dorman, "Long-term cosmic ray intensity variation and part of global climate change, controlled by solar activity through cosmic rays", *Advances in Space Research*, vol. 37 (8), 2006, pp 1621-1628
[2] M. Christl, A. Magnini, S. Holzkamper, C. Spotl., "Evidence for a link between the flux of galactic cosmic rays and Earth's climate during the past 200,000 years", *Journal of Atmospheric and Solar-Terrestrial Physics vol.* 66 (3-4), 2004, pp 313-322
[3] H. Svensmark, E. Friis-Christensen, "Variation of cosmic ray flux and global cloud coverage—a missing link in solar-climate relationships", *Journal of Atmospheric and Solar-Terrestrial Physics,* vol. 59(11), 1997, pp 1225-1232
[4] N. Marsh, H. Svensmark, "Cosmic rays, cloud and climate", *Space Science Review*, vol. 94, 2000, pp. 215-230
[5] H. Svensmark, "Influence of Cosmic Rays on Earth's Climate", *Physical Review Letters,* vol. 81, 2008, pp.5027-5030
[6] I. Usoskin, N. Marsh, G. Kovaltsov et al., "Latidudinal dependence of low cloud amount on cosmic ray induced ionization", *Geophysical Research Letters* vol. 110, 2005, L16109
[7] Z. Begum, "Cosmic ray–aerosol–cloud interactions in the atmospheric environment: Theoretical aspects", *Journal of Quantitative Spectroscopy and Radiative Transfer*, vol. 102(2), 2006, pp. 257-260
[8] R. Harrison, K. Aplin, "Atmospheric condensation nuclei formation and high-energy radiation", *Journal of Atmospheric and Solar-Terrestrial Physics*, vol. 63(17), 2001, pp. 1811-1819
[9] A. Mishev, J. Stamenov, "Present status and further possibilities for space weather studies at BEO Moussala", *Journal of Atmospheric and Solar-Terrestrial Physics "*, vol. 70(2-4), 2008, pp. 680-685
[10] A. Hillas, "Differences between gamma-ray and hadronic showers", *Space Science Review*, vol. 75, 1996, pp. 17-30
[11] K. Bernlohr, "Impact of atmospheric parameters on the atmospheric Cherenkov technique", *Astroparticle Physics*, vol. 12, 2000, pp. 255-268
[12] A. Ivanov, S. Knurenko, I. Sleptsov, "Measuring extensive air showers with Cherenkov light detectors of the Yakutsk array: the energy spectrum of cosmic rays", *New Journal of Physics*, vol. 11, 2009, doi:10.1088/1367-2630/11/6/065008
[13] D. Heck, J. Knapp, J.N. Capdevielle, G. Schatz, T. Thouw, "CORSIKA: A Monte Carlo Code to Simulate Extensive Air Showers". *Report FZKA 6019* Forschungszentrum Karlsruhe, 1997
[14] A. Fasso et al., "The physics models of FLUKA: status and recent developments", Computing in High Energy and Nuclear Physics 2003 Conference (CHEP2003), La Jolla, CA, USA, March 24-28, 2003, (paper MOMT005), eConf C0303241
[15] S. Ostapchenko, "QGSJET-II: towards reliable description of very high energy hadronic interactions", *Nuclear Physics B-Proc. Suppl.*, vol. 151(1), 2006, pp. 143-146
[16] V. Roldugin, B. Tinsley, "Atmospheric transparency changes associated with solar wind-induced atmospheric electricity variations", *Journal of Atmospheric and Solar-Terrestrial Physics*, vol. 66(13-14), 2004, pp. 1143-1149
[17] M. Pudovkin, S. Veretenenko, "Variations of the cosmic rays as one of the possible links between the solar activity and the lower atmosphere", *Advances in Space Research* vol. 17(11), 1996, pp. 161-164
[18] M. Pudovkin, S. V. Veretenenko, R. Pellinen, E. Kyro, "Cosmic ray variation effects in the temperature of the high-latitudinal atmosphere", *Advances in Space Research* vol. 17(11), 1996, pp. 165-168
[19] M. Voiculescu, I. Usoskin, K. Mursula, "Different response of clouds at the solar input", *Geophysical Research Letters* vol. 33, 2006, L21802
[20] J. Kazil, E. Lovejoy, M. Barth, K. O"Brien, "Aerosol nucleation over oceans and the role of galactic cosmic rays", *Atmos. Chem. Phys.* Vol. 6, 2006, pp. 4905-4924
[21] E. Pallé, C.J. Butler, K. O'Brien, "The possible connection between ionization in the atmosphere by cosmic rays and low level clouds", *Journal of Atmospheric and Solar-Terrestrial Physics*, vol. 66(18), 2004, pp. 1779-1790
[22] A. Mishev, I. Anguelov ,J. Stamenov, "Simulations and measurements of Atmospheric Cherenkov light, neutron and muon cosmic ray flux at Basic Environmental Observatory Moussala for space weather studies", *JINST 2 P04002, 2007,* doi:10.1088/1748-0221/2/04/P04002
[23] A. Mishev, J. Stamenov, I. Anguelov, E. Malamova, "About the possibility for estimation of the atmospheric transparency on the basis of EAS Cherenkov light registration", *Proc. of 30th ICRC* Merida, Mexico 2007, Vol. 5, Pp. 845-848
[24] A. Rapti, "Atmospheric transparency, atmospheric turbidity and climatic parameters", *Solar Eenery*, vol. 69(2), 2000, 99-111
[25] C. Perry, "Evidence for a physical linkage between galactic cosmic rays and regional climate time series", *Advances in Space Research*, Vol. 40(3), 2007, pp. 353-364
[26] M. Pudovkin, S. V. Veretenenko, "Cloudiness decreases associated with Forbush-decreases of galactic cosmic rays", *Journal of Atmospheric and Terrestrial Physics*, Vol. 57(11), 1995, pp. 1349-1355
[27] I. Usoskin, G. Kovaltsov, "Cosmic rays and climate of the Earth: Possible connection", *Comptes Rendus Geosciences*, Vol. 340(7), 2008, pp. 441-450
[28] L. Desorgher, E. Fluckiger, M. Gurtner, M. Moser, R. Butikofer, "Atmocosmics: A GEANT4 code for computing the interaction of cosmic rays with the Earth's atmosphere", *Int. J. Mod. Phys. A*, 20 (29), 2005, pp. 6802– 6804
[29] I. Usoskin, G. Kovaltsov, "Cosmic ray induced ionization in the atmosphere: Full modeling and practical applications", *J. Geophys. Res.*, Vol. 111, 2006, D21206
[30] I. Usoskin., L. Desorgher, P. Velinov et al., "Ionization of the Earth's atmosphere by galactic and solar cosmic rays", *Acta Geophysica,,* Vol. 57(1), 2009, pp. 88-101
[31] P.I.Y. Velinov, A. Mishev, L. Mateev, "Model for induced ionization by galactic cosmic rays in the Earth atmosphere and ionosphere", *Advances in Space Research*, Vol. 44(9), 2009, pp. 1002-1007
[32] K. Carslaw, R. Harrison R, J. Kirkby, "Cosmic Rays, Clouds and Climate", *Science*, vol. 298, 2002, pp. 1732-1737
[33] F. L. Eisele, R. Lovejoy E, E. Kosciuch et al., "Negative atmospheric ions and their potential role in ion-induced nucleation", *J. Geophys. Res*, vol. 111, 2005, D04305, doi:10.1029/2005JD006568
[34] J. Curtius, E. Lovejoy, K. Froyd, "Atmospheric Ion-induced Aerosol Nu-cleation", *Space Science Reviews,* vol. 125(1-4), 2006, pp. 159-167
[35] B. Tinsley, R. Rohrbaugh, M. Hei, "Electroscavenging in clouds with broad droplet size distributions and weak electrifcation", *Atmospheric Research*, vol. 59-60, 2001, pp. 115-135.
[36] S. Tripathi and R. Harrison, "Enhancement of contact nucleation by scavenging of charged aerosol particles", *Atmospheric Research*, vol. 62(1-2), 2002, pp. 57- 70.
[37] B. Tinsley, L. Zhou, A. Plemmons, "Changes in scavenging of particles by droplets due to weak electrifcation in clouds", *Atmospheric Research*, vol. 79(3-4), 2006, pp. 266-295.